\documentstyle[preprint,aps,epsfig,amsbsy,floats]{revtex}
 \tightenlines

\begin{document}
\newcommand{\beq}{\begin{equation}}
\newcommand{\eeq}{\end{equation}}
\newcommand{\beqa}{\begin{eqnarray}}
\newcommand{\eeqa}{\end{eqnarray}}
\newcommand{\sr}{\sqrt}
\newcommand{\fr}{\frac}
\newcommand{\mn}{\mu \nu}
\newcommand{\G}{\Gamma}

\draft \preprint{hep-th/0402018,~ INJE-TP-04-01}
\title{Noncommutative spacetime effect on the slow-roll period of inflation}
\author{ Hungsoo Kim, Gil Sang Lee, and  Yun Soo Myung\footnote{E-mail address:
ysmyung@physics.inje.ac.kr}}
\address{
Relativity Research Center and School of Computer Aided Science\\
Inje University, Gimhae 621-749, Korea} \maketitle

\begin{abstract}
We study how the noncommutative spacetime affects on inflation.
First we obtain  the noncommutative power spectrum of the
 curvature perturbations produced during inflation in the slow-roll
 approximation. This is the explicit $k$-dependent power spectrum up to first order
 in slow-roll parameters $\epsilon_1,~\delta_1$ including the noncommutative parameter $\mu$.
In order to test the role of $\mu$ further, we calculate the
noncommutative power spectrum  using the slow-roll expansion. We
find  corrections which arise from  the change of pivot scale
 and  a noncommutative parameter  with $\mu\not=$
constant. It turns out that the noncommutative parameter $\mu$
could  be considered as a zeroth order slow-roll parameter and the
noncommutative spacetime effect suppresses the power spectrum.

\end{abstract}

\thispagestyle{empty}
\setcounter{page}{0}
\newpage
\setcounter{page}{1}

\section{Introduction}
Sting  theory as a candidate for the theory of everything can say
something  about cosmology\cite{Bran1}. Focusing on a universal
property of string theory, it is very interesting to study its
connection to cosmology. The universal property which we wish to
choose here is a new uncertainty relation of $ \triangle t_p
\triangle x_p \ge l^2_s$ where $l_{s}$ is the string length
scale\cite{Yone}. This implies that spacetime is noncommutative.
It is compared to a stringy uncertainty relation of  $ \triangle
x_p \triangle p \ge 1+ l^2_s \triangle p^2$. The former is
considered as a universal property for strings as well as
D-branes, whereas the latter is suitable only for strings.
Spacetime noncommutativity does not affect the evolution of the
homogeneous background. However, this leads to a coupling between
the fluctuations generated in inflation and the flat background of
Friedmann-Robertson-Walker (FRW) space\cite{BH}. Usually the
coupling appears to be nonlocal in time.

On the other hand, it is generally accepted that curvature
perturbations produced during inflation are considered  to be
 the origin of  inhomogeneities necessary for
explaining  galaxy formation and other large-scale structure. The
first year results of WMAP put forward more constraints on
cosmological models and confirm the emerging standard model of
cosmology, a flat $\Lambda$-dominated universe seeded by
scale-invariant adiabatic gaussian fluctuations\cite{Wmap1}. In
other words, these results coincide with  predictions of the
inflationary scenario with an inflaton.  Also WMAP brings about
some new intriguing results: a running spectral index of scalar
metric perturbations and an anomalously low quadrupole of the CMB
power spectrum\cite{Wmap2}. If inflation is affected by physics at
scale close to string scale $l_s$, one expects that the spacetime
uncertainty must be encoded in the CMB power spectrum and running
spectral index. For example, the noncommutative power-law
inflation may produce a large running spectral index to fit the
data of WMAP\cite{HM1,TMB,HM2}.

Recently the noncommutative power spectrum, spectral index, and
running spectral index of the curvature perturbations produced
during inflation have been calculated  with  the slow-roll
parameters $\epsilon_1$ and $\delta_n$ and noncommutative
parameter $\mu$\cite{HM3}.

In this work we  examine whether or not the noncommutative
parameter $\mu$  is considered as a slow-roll parameter.  For this
purpose, first we use the slow-roll approximation to obtain the
explicit $k$-dependent power spectrum up to first order. Here $k$
is the index for comoving Fourier modes. Then we introduce the
slow-roll expansion to find the noncommutative power spectrum up
to first order. It turns out that the noncommutative parameter
$\mu$ is   considered as a zeroth order slow-roll parameter.

\section{Commutative Case}

Our starting point is the effective
action during inflation,
\begin{equation}
S = \int \left[ -\frac{M^{2}_{Pl}}{2}R + \frac{1}{2}(\partial_\mu
\phi)^2 - V(\phi) \right] \sqrt{-g} \ d^4 x,
\end{equation}
where $M^{2}_{P}$ is the reduced Planck mass defined by
$M_{P}=(8\pi G)^{1/2}$. For simplicity we choose $M^{2}_{P}=1$.
 The scalar  metric perturbation to the homogeneous, isotropic
background  is expressed  in longitudinal gauge as \cite{bardeen}
\begin{equation}
\label{con-p}
 ds^2_{con-p} = a^2(\eta) \left\{(1+2A) d\eta^2
-(1+2\psi)d{\bf x}\cdot d{\bf x} \right\},
\end{equation}
where the conformal time $\eta$ is given by $d\eta = dt / a $. We
get a relation of $\psi=A$ because the stress-energy tensor does
not have any off-diagonal component. It is convenient to express
the density perturbation in terms of the curvature perturbation
$R_c$ of comoving hypersurfaces given by~\cite{lukash}
\begin{equation}
R_c = \psi-\frac{H}{ \dot \phi} \delta \phi
\end{equation}
during inflation, where $\delta\phi$ is the perturbation in
inflaton: $\phi({\bf x},\eta)=\phi(\eta)+\delta\phi({\bf
x},\eta)$. The overdot is derivative with respect to a comoving
time $t$ defined in the flat FRW line element:
$ds^2_{FRW}=dt^2-a(t)^2 d{\bf x}\cdot d{\bf x}$.  The power
spectrum $P_{R_c}(k)$ of curvature perturbation  is defined by
\begin{equation}
\langle R_c({\bf x},\eta), R_c({\bf y},\eta)\rangle = \int
\frac{dk}{k} \frac{\sin(k|{\bf x}-{\bf y}|)}{k|{\bf x}-{\bf y}|}
P_{R_c}.
\end{equation}

Defining
\begin{equation}
z \equiv \frac{a\dot \phi}{H} \ \ \ \mbox{and}\ \ \ \varphi \equiv
a\left(\delta\phi-\frac{\dot\phi}{H} \psi\right) = -z R_c,
\end{equation}
the bilinear action for curvature  perturbation is \cite{mukhanov}
\begin{equation}
\label{bilinear} S = \int \frac{1}{2} \left[
\left(\frac{\partial\varphi}{\partial\eta}\right)^2 -
\left(\nabla\varphi\right)^2 +
\left(\frac{1}{z}\frac{d^2z}{d\eta^2}\right)\varphi^2 \right]
d\eta\,d^3{\bf x}.
\end{equation}
At this stage we wish to note that $z$ encodes information about
inflation.  Because the background is spatially flat, we can
expand all perturbed fields in terms of Fourier modes as $ f({\bf
x},\eta)=\int \fr{d^3{\bf k}}{(2 \pi)^{3/2}} f_{{\bf k}}(\eta)
e^{i{\bf k}\cdot {\bf x}}$. In second quantization\footnote{For
example, the orthogonal eigenmodes expansion of $ \varphi({\bf
x},\eta)$ is given by $\varphi({\bf x},\eta)=\int \fr{d^3{\bf
k}}{(2 \pi)^{3/2}}\Big( b({\bf k})u_k(\eta) e^{i{\bf k}\cdot {\bf
x}}+b^\dagger({\bf k})u^*_k(\eta) e^{-i{\bf k}\cdot {\bf
x}}\Big)$. Here the annihilation and creation operators $b({\bf
k}),b^\dagger({\bf k})$ satisfy the usual operator algebra :
$\left[ b({\bf k}),b^\dagger({\bf q})\right]=\delta^{(3)}({\bf
k}-{\bf q})$.}, these modes are given by $f_{{\bf k}}(\eta)=b({\bf
k})f_k(\eta) +b^\dagger(-{\bf k})f^*_k(\eta)$. This
quantum-to-classical behavior is a great success for the theory.
If it had failed,  prediction for the power spectrum  would have
had nothing to do with reality. Using the Fourier transform of
$\varphi$ and second quantization, each mode of curvature
perturbation satisfies the Schr\"odinger-type equation
\begin{equation}
\label{eqs7} \frac{d^2\varphi_k(\eta)}{d\eta^2} +
\left(k^2-\frac{1}{z}\frac{d^2z}{d\eta^2}\right)\varphi_k(\eta) =
0,~~k=|{\bf k}|,
\end{equation}
where  $\varphi_k$ depends on the norm of ${\bf k}$ only because
we work in an isotropic background. In general its asymptotic
solutions are obtained as
\begin{equation}\label{bc}
\varphi_k \longrightarrow \left\{
\begin{array}{l l l}
\frac{1}{\sqrt{2k}}e^{-ik\eta} & \mbox{as} & -k\eta \rightarrow \infty \\
A_k z & \mbox{as} & -k\eta \rightarrow 0.
\end{array} \right.
\end{equation}
 The first solution corresponds to the flat space vacuum on
scales much smaller than the Hubble distance (sub-horizon scale),
and the second is the growing mode on scales much larger than the
Hubble distance (super-horizon scale). Using  other definition of
the power spectrum : $P_{R_c}(k)\delta^{(3)}({\bf k}-{\bf
q})=\fr{k^3}{2\pi^2}<R_{c{\bf k}}(\eta)R^{\dagger}_{c{\bf
q}}(\eta)>$, one finds
\begin{equation}\label{gps}
P_{R_c}(k) = \left(\frac{k^3}{2\pi^2}\right)
\lim_{-k\eta\rightarrow0}\left|\frac{\varphi_k}{z}\right|^2 =
\frac{k^3}{2\pi^2}|A_k|^2.
\end{equation}
Now let us  introduce slow-roll parameters
\begin{equation}
\epsilon_1 = -\frac{\dot H}{H^2} =
\frac{1}{2}\left(\frac{\dot{\phi}}{H}\right)^2,~~\delta_n \equiv
\frac{1}{H^n\dot{\phi}}\frac{d^{n+1}\phi}{dt^{n+1}},
\end{equation}
which we assume to satisfy $\epsilon_1 < \xi$ and $|\delta_n| <
\xi^n$ for a small perturbation parameter $\xi$. The subscript
denotes the order in the slow-roll expansion. Then one finds
relations up to first order
 \beq
\frac{1}{z}\frac{d^2z}{d\eta^2} \simeq
2(aH)^2\Big(1+\epsilon_1+\fr{3}{2}\delta_1\Big),~~aH\simeq
-\fr{1}{\eta}(1+\epsilon_1). \eeq Ignoring the slow-roll
corrections, equation (\ref{eqs7}) is given by \beq
\frac{d^2\varphi_k}{d\eta^2} +
\left(k^2-\frac{2}{\eta^2}\right)\varphi_k = 0.\eeq Eigenmode
solution to this equation is given by
$\varphi_k=\frac{1}{\sqrt{2k}}\Big(1-\fr{i}{k\eta}\Big)e^{-ik\eta}$.
The power spectrum  to zeroth order in the slow-roll expansion is
then given by\cite{LL}
\begin{equation}
\label{sol0} P^{0th}_{R_c}(k)
=\left.\left(\frac{H}{2\pi}\right)^2\left(\frac{H}{\dot\phi}\right)^2
\right|_{k=k_*},
\end{equation}
where we use the relation \beq\label{relation} <\fr{\varphi_{{\bf
k}}}{a}\fr{\varphi^{\dagger}_{{\bf q}}}{a}>=
\frac{2\pi^2}{k^3}\left(\frac{H}{2\pi}\right)^2\delta^{(3)}({\bf
k}-{\bf q}). \eeq In commutative spacetime, the power spectrum is
usually evaluated at horizon crossing
 which is the moment of the time when $k_*=aH$. The scale $k_*$ is called
 the pivot scale.  The standard slow-roll approximation (Bessel approximation) is
valid only when slow-roll parameters are considered to be small
and nearly constant\cite{SL}. This gives us  the power spectrum to
first order in the slow-roll expansion. Using \beq
\label{zz}\frac{1}{z}\frac{d^2z}{d\eta^2} \simeq
\fr{(2+6\epsilon_1+3\delta_1)}{\eta^2}=\fr{\nu^2-\fr{1}{4}}{\eta^2},\eeq
 with $\nu=\fr{3}{2}+2\epsilon_1+\delta_1$, we can
calculate the power spectrum  to first order \beq
\label{ps1st}P^{1st}_{R_c}(k) = \frac{H^4}{(2\pi)^2\dot\phi^2}
\left\{ 1 -2\epsilon_1 +2\alpha(2\epsilon_1+\delta_1) \right\}
\eeq where $\alpha=2-\ln2-\gamma=0.729637$ with $\gamma=0.577216$
(Euler-Mascheroni constant). Here all quantities are evaluated at
$k=k_*=aH$.  The slow-roll approximation cannot be pushed beyond
the first order. In order to calculate the power spectrum up to
second order, one should use the other method called slow-roll
expansion based on Green functions perturbative
calculation\cite{SG}. In this approach an important point is that
the slow-roll parameters satisfy the relations: \beq
\label{slow-di}\dot \epsilon_1
=2H(\epsilon_1^2+\epsilon_1\delta_1),~~\dot{\delta}_1=H(\epsilon_1\delta_1-\delta^2_1+\delta_2),~~
\dot \delta_2=H(2\epsilon_1\delta_2-\delta_1\delta_2+\delta_3)
\eeq which means that the derivative of slow-roll parameters with
respect to time increases their  order by one in the slow-roll
expansion. In other words, all parameters of
$\epsilon_1,~\delta_n$ are not constant but slowly varying because
$H$ is nearly constant and $\epsilon_1<\xi,~|\delta_n|<\xi^n$ for
a small parameter $\xi<1$ during inflation. Using Eq.(\ref{gps}),
the power spectrum up to the second order is given by

\begin{eqnarray}
 \label{2ndps}
P^{2nd}_{R_c}(k) & = & \frac{H^4}{(2\pi)^2\dot\phi^2} \left\{ 1
-2\epsilon_1  + 2\alpha(2\epsilon_1+\delta_1)+
\left(4\alpha^2-23+\frac{7\pi^2}{3}\right)\epsilon_1^2 \right.\\
&& \left. +
\left(3\alpha^2+2\alpha-22+\frac{29\pi^2}{12}\right)\epsilon_1\delta_1
+ \left(3\alpha^2-4+\frac{5\pi^2}{12}\right)\delta_1^2 +
\left(-\alpha^2+\frac{\pi^2}{12}\right)\delta_2 \right\},
\nonumber
\end{eqnarray}
where all quantities are  evaluated at horizon crossing of
$k=k_*$. Comparing Eq.(\ref{ps1st}) with Eq.(\ref{2ndps}) leads to
the same power spectrum up to first order.

\section{Noncommutative Case}

For convenience we introduce another time coordinate
$\tau$ to represent the noncommutative spacetime. Then the
perturbed metric in Eq.(\ref{con-p}) can be rewritten as
\begin{equation}
ds^2_{non-p} = a^{-2}(\tau)(1+2A) d\tau^2 - a^2(\tau)(1+2\psi)
d{\bf x}\cdot d{\bf x}.
\end{equation}
The spacetime uncertainty relation of $ \triangle t_p \triangle
x_p \ge l^2_s$ becomes
\begin{equation}\label{non-c}
\triangle \tau \triangle x \ge l^2_s \eeq for a cosmological
purpose. We propose the transition to noncommutative spacetime
obeying Eq.(\ref{non-c}) by taking the operator appearing in the
bilinear action in Eq.(\ref{bilinear}) and replacing all
multiplications by $*$-products\cite{BH}.  Using the Fourier
transform of $\tilde{\varphi}$, the equation of motion becomes
\begin{equation}
\label{eqsn} \frac{d^2\tilde{\varphi}_k}{d\tilde{\eta}^2} +
\left(k^2-\frac{1}{z_k}\frac{d^2z_k}{d\tilde{\eta}^2}\right)\tilde{\varphi}_k
= 0.
\end{equation}
Here $z_k$ is some smeared version of $z$ or $a$ over a range of
time of characteristic scale $\triangle \tau =l_s^2k$ defined by
\beq \label{zzz} z^2_k(\tilde{\eta})=z^2y^2_k(\tilde{\eta}),
~~y^2_k=\sqrt{\beta^+_k\beta^-_k},~~\fr{d\tilde{\eta}}{d\tau}
=\sqrt{\fr{\beta^-_k}{\beta^+_k}}, \eeq where \beq \label{betak}
\beta^{\pm}_k=\fr{1}{2} \left[a^{\pm 2}(\tau+l^2_sk)+a^{\pm
2}(\tau-l^2_sk)\right]. \eeq  Here we define a modified conformal
time $\tilde{\eta}$.  This representation has the advantage of
preserving both spatial translational and rotational symmetry of
the flat FRW metric, in compared to constructions based on the
conventional noncommutative relations: $
\left[x^\mu,x^\nu\right]=i \theta^{\mn}$\cite{FKM}. Actually
spacetime noncommutativity does not affect the evolution of the
homogeneous background. However, this leads to a coupling between
the fluctuations generated in inflation and the flat background of
FRW space. Usually the coupling appears to be nonlocal in time as
is shown in Eq.(\ref{zzz}).  If one does not require the
uncertainty relation in Eq.(\ref{non-c}), one finds easily
commutative relations that $y_k \to 1, ~z_k \to z, \tilde{\eta}\to
\eta$.

Our task is to solve Eq.~(\ref{eqsn}). Relevant parameters are
$\epsilon_1,~\delta_n$  as well as   a small noncommutative
parameter $\mu(k,t)$~\footnote{If one considers $\mu(t)$, then one
finds $\dot \mu=-2 H\mu(1+\epsilon_1)$ which is similar to
$\dot{X}=-2HX(1-\epsilon_1)$ with $X=(aH)^{-2}$ in Ref.\cite{ZS}.
This leads to a wrong interpretation for the noncommutative
parameter $\mu$ as a zeroth order slow-roll parameter because
$\dot{\mu}(t)$ has two terms : one is a zeroth-order and the other
is a first order in the slow-roll expansion. We thank Q. Huang for
pointing out it.} defined as
\begin{equation}
\mu=\Big(\fr{kH}{aM^2_s}\Big)^2
\end{equation}
which satisfies \beq \label{mudi}\dot \mu=-4 H\mu\epsilon_1.\eeq
 Here
$M_s=1/l_s$. At the first time it seems that  difference exists
between $\epsilon_1,~\delta_n$ and $\mu$.  By definition,
slow-roll parameters ($\epsilon_1,~\delta_n<<1$) do not involve
$a$ which is rapidly increasing during inflation. We note that $H$
is nearly constant during inflation. $\mu$ does not contains any
derivative with respect to time but it contains $a$, in contrast
to $\epsilon_1,~\delta_n$. Even though $\mu$ has a different
property, we insist from Eq.(\ref{mudi}) that $\mu$ can play the
role of a small parameter. This is because  assuming that $\mu$ is
zeroth order, derivative of $\mu$ becomes first order in the
slow-roll expansion. Hence we take $\mu$ as a zeroth order of
slow-roll parameter which describes the noncommutative spacetime
at the period of inflation.

\subsection{Slow-roll approximation}
In this section we use the slow-roll approximation which means
that we take  $\epsilon_1,~\delta_1,~\mu$ to be constant in
calculation of the noncommutative power spectrum. For this
purpose, we obtain relations up to first order\cite{HM3}
 \beq \label{nonzzz}
\frac{1}{z_k}\frac{d^2z_k}{d\tilde{\eta}^2}\simeq
2(aH)^2\Big(1+\epsilon_1+\fr{3}{2}\delta_1 -2\mu\Big) \eeq and
\beq \label{nonaH}aH\simeq
-\fr{1}{\tilde{\eta}}(1+\epsilon_1+\mu).
 \eeq
 Then Eq.(\ref{eqsn}) takes the form
 \beq
\frac{d^2\tilde{\varphi}_k}{d\tilde{\eta}^2} +
\left(k^2-\frac{(\nu^2-\fr{1}{4})}{\tilde{\eta}^2}\right)\tilde{\varphi}_k
= 0\eeq with $\nu=\fr{3}{2}+2\epsilon_1+\delta_1$. We note here
that this equation takes the same form as in Eq.(\ref{eqs7}) with
Eq.(\ref{zz}) except replacing $\eta,~\varphi_k$ by
$\tilde{\eta},~\tilde{\varphi}_k$. The asymptotic solution to
Eq.(\ref{eqsn}) in the limit of $-k\tilde{\eta} \to \infty$ takes
the form
\begin{equation}\label{nbc}
\tilde{\varphi}_k =\frac{1}{\sqrt{2k}}e^{-ik\tilde{\eta}}.
\end{equation}
In the limit of $-k\tilde{\eta} \to 0$, one finds asymptotic form
of the Hankel function $H^{(1)}_\nu(-k\tilde{\eta})$ \cite{SL}\beq
\label{nonphi}\tilde{\varphi}_k \simeq
e^{i(\nu-\fr{1}{2})}2^{\nu-\fr{3}{2}}\fr{\Gamma(\nu)}{\Gamma(\fr{3}{2})}
\fr{1}{\sqrt{2k}}(-k\tilde{\eta})^{\fr{1}{2}-\nu}. \eeq
Furthermore from Eqs.(\ref{zzz}) and (\ref{betak}), we have an
expression \beq \label{yk}y_k\simeq 1+\mu. \eeq The Fourier
transform of curvature perturbation is given by
$\tilde{R}_k=-\tilde{\varphi}_k(\tilde{\eta})/z_k$.
 Then the  noncommutative power spectrum is defined  by
\begin{equation}\label{nps}
\widetilde{P}_{R_c}(k) = \left(\frac{k^3}{2\pi^2}\right)
\lim_{-k\eta\rightarrow0}\left|\frac{\tilde{\varphi}_k}{z_k}\right|^2.
\end{equation}
Substituting equations (\ref{nonaH}), (\ref{nonphi}) and
(\ref{yk}) into Eq.(\ref{nps}), one finds
 \beq
\widetilde{P}_{R_c}(k)
=\frac{H^4}{(2\pi)^2\dot\phi^2}\Big[2^{\nu-\fr{3}{2}}\fr{\Gamma(\nu)}{\Gamma(\fr{3}{2})}\Big]^2
\Big(\fr{k}{aH}\Big)^{-2(2\epsilon_1+\delta_1)}
\fr{1}{(1+\epsilon_1+\mu)^{2(1+2\epsilon_1+\delta_1)}(1+\mu)^{2}}.
\eeq Making use of  the Taylor expansions up to first order as\beq
2^{\nu-\fr{3}{2}}\fr{\Gamma(\nu)}{\Gamma(\fr{3}{2})} \simeq
1+2\alpha(2\epsilon_1+\delta_1),~~e^{-2(2\epsilon_1+\delta_1)\ln\Big(\fr{k}{k_*}\Big)}\simeq
1- 2(2\epsilon_1+\delta_1)\ln\Big(\fr{k}{k_*}\Big),\eeq we have
 \beq \label{nonps1} \tilde{P}^{1st}_{R_c}(k)
= \frac{H^4}{(2\pi)^2\dot\phi^2} \left\{ 1 -2\epsilon_1 -4\mu +2
\left(\alpha-\ln\Big(\fr{k}{k_*}\Big)\right)(2\epsilon_1+\delta_1)
\right\}  \eeq with $k_*=aH$. In the limit of $\mu \to 0$,
$\widetilde{P}^{1st}_{R_c}(k)$ reduces to the commutative power
spectrum\cite{MS1,STW}. In the noncommutative spacetime approach
the horizon crossing occurs at
$k^2=\frac{1}{z_k}\frac{d^2z_k}{d\tilde{\eta}^2}$\cite{BH}. Hence
from Eq.(\ref{nonzzz}) we use the other pivot scale $k_{nhc}\simeq
\sqrt{2}k_*$. As a result, we obtain the noncommutative power
spectrum up to first order as
 \beq \label{nonps11}\tilde{P}^{1st}_{R_c}(k)
= \frac{H^4}{(2\pi)^2\dot\phi^2} \left\{ 1 -2\epsilon_1 -4\mu+
2\alpha_*(2\epsilon_1+\delta_1) \right\} \eeq with
$\alpha_*=\alpha-\fr{\ln2}{2}$. Here the right hand side is
evaluated at $k=k_{nhc}$. Comparing it with Eq.(\ref{ps1st}) when
$\mu=0$, the change of pivot scale from $k=k_*$ to $k=k_{nhc}$
amounts to replacing $\alpha$ by $\alpha_*$ in the first-order
calculation\cite{STW}.

\subsection{Slow-roll expansion}
The slow-roll approximation could not be considered as a general
approach to calculate the power spectrum. In order to calculate
the power spectrum even for up to first order correctly, one
should use the slow-roll expansion based on Green's function
technique. The key step is to introduce a variable nature of the
noncommutative parameter, which means that $\mu$ satisfies
Eq.(\ref{mudi}). However this expansion is useful for deriving the
power spectrum at $k=k_*$ but not $k=k_{nhc}$ and thus it works
well for  commutative case and higher order case.  We use the
slow-roll expansion at $k=k_*$ to calculate the noncommutative
power spectrum up to first order. Then, accepting the rule that
the change of  pivot scale from $k=k_*$ to $k=k_{nhc}$ amounts to
replacing $\alpha$ by $\alpha_*$, our calculation  provides the
correct result that will be derived from the slow-roll expansion
at $k=k_{nhc}$.

Using notations of $y = \sqrt{2k}\, \tilde{\varphi}_k$ and
$\tilde{x} = -k\tilde{\eta}$, we can express Eq.(\ref{eqsn}) as
\begin{equation} \label{eqsnn}
\frac{d^2y}{d\tilde{x}^2} +
\left(1-\frac{1}{z_k}\frac{d^2z_k}{d\tilde{x}^2}\right)y=0.
\end{equation}In general its asymptotic solutions
are given by
\begin{equation}
\label{z-sol} y \longrightarrow \left\{
\begin{array}{l l l}
e^{i\tilde{x}} & \mbox{as} & \tilde{x} \rightarrow \infty \\ \\
\sqrt{2k}\,\tilde{A}_k z_k & \mbox{as} & \tilde{x} \rightarrow 0.
\end{array}
\right.
\end{equation}

We solve Eq.~(\ref{eqsnn}) with the boundary condition
Eq.~(\ref{z-sol}) to eventually calculate $\tilde{A}_k$. Now we
can choose the ansatz that $z_k$ takes the form
\begin{equation}
z_k = \frac{1}{\tilde{x}}\widetilde{f}(\ln \tilde{x}).
\end{equation}
Then we have
\begin{equation}
\frac{1}{z_k}\frac{d^2z_k}{d\tilde{x}^2} = \frac{2}{\tilde{x}^2} +
\frac{1}{\tilde{x}^2}\widetilde{g}(\ln \tilde{x}),
\end{equation}
where
\begin{equation}\label{gf}
\widetilde{g}=\frac{-3\widetilde{f}'+\widetilde{f}''}{\widetilde{f}},
\end{equation}
and the equation of motion is
\begin{equation}
\label{exactsol} \frac{d^2y}{d\tilde{x}^2} +
\left(1-\frac{2}{\tilde{x}^2}\right) y =
\frac{1}{\tilde{x}^2}\widetilde{g}(\ln \tilde{x}) y.
\end{equation}
The homogeneous solution with correct asymptotic behavior at
$\tilde{x} \rightarrow\infty$ is
\begin{equation}
\label{0sol} y_0(\tilde{x}) = \left(1 +
\frac{i}{\tilde{x}}\right)e^{i\tilde{x}}.
\end{equation}
Using Green's function technique, Eq.(\ref{exactsol}) with the
boundary condition Eq.(\ref{z-sol}) can be written as the integral
equation
\begin{equation}
\label{sol} y(\tilde{x}) = y_0(\tilde{x}) +
\frac{i}{2}\int_{\tilde{x}}^{\infty}du \ \frac{1}{u^2} \
\widetilde{g}(\ln u) \ y(u)
\left[y_0^*(u)y_0(\tilde{x})-y_0^*(\tilde{x})y_0(u)\right].
\end{equation}

We are now in a position to solve Eq.~(\ref{sol}) perturbatively
using the slow-roll expansion. Introducing
\begin{equation}\label{zexp}
\tilde{x}z_k = \widetilde{f}(\ln \tilde{x}) = \sum_{n=0}^{\infty}
\frac{\widetilde{f}_n}{n!}(\ln \tilde{x})^n,
\end{equation}
 $\widetilde{f}_n/\widetilde{f}_0$ is of order $n$ in the slow-roll expansion. This
expansion is useful for $\exp(-1/\xi) \ll \tilde{x} \ll
\exp(1/\xi)$ and for extracting information at $\tilde{x}=1$.

Considering a relation up to first order in the slow-roll
expansion and in $\mu$ as
\begin{equation}
\tilde{x} = -k\tilde{\eta} = -k\int d\tau
\left(\fr{\beta^-_k}{\beta^+_k}\right)^{1/2} \simeq
\frac{k}{aH}\left\{1+ \epsilon_1+\mu(1-2\epsilon_1)
 \right\},
\end{equation}
we can express the expansion coefficients $\tilde{f}_n$ in terms
of $\epsilon_1$, $\delta_n$, and $\mu$ evaluated at $k=aH$. In
deriving the above expression,  we use Eq.(\ref{mudi}). From
Eq.(\ref{zexp}) we  obtain
\begin{eqnarray}
\tilde{f}_1  \simeq
\left.-\frac{k\dot\phi}{H^2}\left\{2\epsilon_1+\delta_1+2\mu(-2\epsilon_1+\delta_1)
                            \right\}
\right|_{k=aH}, \\
\tilde{f}_0  \simeq  \left.
\frac{k\dot\phi}{H^2}\left\{1+\epsilon_1+\mu(2
+\epsilon_1+\delta_1)
                     \right\}
\right|_{k=aH},\\
\frac{1}{\tilde{f}_0}\simeq \left.\frac{H^2}{k\dot\phi}\left\{
1-\epsilon_1-\mu(2-3\epsilon_1
                          +\delta_1)\right\}
                            \right|_{k=aH},\\
\frac{\tilde{f}_1}{\tilde{f}_0}\simeq \left.
\left(-2\epsilon_1-\delta_1+8\mu\epsilon_1\right)
                                     \right|_{k=aH}.
\end{eqnarray}
Further Eqs.(\ref{gf}) and~(\ref{zexp}) give
\begin{equation}\label{gexp}
\tilde{g}(\ln \tilde{x}) = \sum_{n=0}^{\infty}
\frac{\tilde{g}_{n+1}}{n!}(\ln \tilde{x})^n,
\end{equation}
where $\tilde{g}_n$ is of order $n$ in the slow-roll expansion
and, up to first order \beq\label{g1} \tilde{g}_1  \simeq
-3\frac{\tilde{f}_1}{\tilde{f}_0}. \eeq Expanding $y$ as
\begin{equation}\label{yexp}
y(\tilde{x}) = \sum_{n=0}^{\infty}y_n\left(\tilde{x}\right),
\end{equation}
where $y_0(\tilde{x})$ is the homogeneous solution in
Eq.(\ref{0sol}), and $y_n(\tilde{x})$ is of order $n$ in the
slow-roll expansion. Following the procedure in commutative
case~\cite{SG}, we solve Eq.~(\ref{sol}) perturbatively by
substituting Eqs.(\ref{gexp}) and~(\ref{yexp}) and equating terms
of the same order. We obtain the asymptotic form for $y$ up to
first-order corrections
\begin{eqnarray}\label{final-y}
y(\tilde{x}) & \rightarrow & i \left\{ 1 + \frac{\tilde{g}_1}{3}
\left[\alpha+\frac{i\pi}{2}\right] \right\} \tilde{x}^{-1} \nonumber\\
&& \mbox{} - \frac{i}{3}  \tilde{g}_1  \tilde{x}^{-1}\ln
\tilde{x}.
\end{eqnarray}
The exact asymptotic form for $y$ in the limit
$\tilde{x}\rightarrow 0$ is given by Eq.~(\ref{z-sol}). Expanding
this perturbatively as in Eq.~(\ref{zexp}) for small
$\xi\ln(1/\tilde{x})$, $\mbox{i.e.}$ for $\tilde{x}$ in the range
$1 \gg \tilde{x} \gg \exp(-1/\xi)$, gives the asymptotic form for
$y$ up to first-order corrections
\begin{equation}\label{asymp}
y(\tilde{x}) \rightarrow \sqrt{2k}\,\tilde{A}_k \tilde{f}_0 \
\tilde{x}^{-1} + \sqrt{2k}\,\tilde{A}_k \tilde{f}_1 \
\tilde{x}^{-1}\ln \tilde{x}.
\end{equation}
Comparing this with Eq.~(\ref{final-y}), the coefficient of
$\tilde{x}^{-1}$ is the desired result because it will give
$\tilde{A}_k$ up to first-order corrections. The coefficient of
$\tilde{x}^{-1}\ln \tilde{x}$ simply give the consistent
asymptotic behavior, that is, proportional to $z_k$. Substituting
Eq.(\ref{g1}) into Eq.~(\ref{final-y}),
 matching the coefficient of $\tilde{x}^{-1}$ with that in
 Eq.(\ref{asymp}),
  the noncommutative power spectrum up to first order is
\begin{equation}\label{amp}
\tilde{P}^{1st}_{R_c}(k) =\frac{k^3}{2\pi^2}|\tilde{A}_k|^2=
\frac{k^2}{(2\pi)^2}\frac{1}{\tilde{f}_0^2} \left[1 -
2\alpha\frac{\tilde{f}_1}{\tilde{f}_0}\right].
\end{equation}
Substituting Eqs.(49)-(50) into Eq.(\ref{amp}) leads to
\begin{eqnarray}
\label{nonps1st} \tilde{P}^{1st}_{R_c}(k) & = &
P^{1st}_{R_c}(k)-\frac{\mu H^4}{(2\pi)^2\dot\phi^2} \left\{
       4+(32\alpha-10)\epsilon_1 + (8\alpha+2)\delta_1
\right\}
\end{eqnarray}
where the commutative contribution appears in Eq.(\ref{ps1st}) and
the right hand side should be evaluated at $k=aH$. Comparing with
the result Eq.(\ref{nonps11}) from the slow-roll approximation, we
find additional terms depending $\mu$. We identify these with  an
effect of choosing Eq.(\ref{mudi}). Using $\fr{d\mu}{d\ln k}=
-4\epsilon_1 \mu,~ \fr{d\epsilon_1}{d\ln k}=
2(\epsilon_1^2+\epsilon_1\delta_1),~\fr{d\delta_1}{d\ln k}=
\epsilon_1\delta_1-\delta_1^2+\delta_2$, the spectral index
defined by
\begin{equation}
\tilde{n}_{s}(k) = 1 + \frac{d \ln \tilde{P}^{1st}_{R_c}}{d \ln k}
\end{equation}
can be easily calculated up to second order
\begin{eqnarray}
\tilde{n}_{s}(k) = & &n_{s}(k) + 16 \mu \epsilon_1 \\ \nonumber
                  +&  & \mu \left\{
                   (32\alpha+12)\epsilon_1^2
                   - (32\alpha-10)\epsilon_1\delta_1
                   + 2\delta_1^2 - 2\delta_2
                   \right\}.
\end{eqnarray}
where the right hand side should be evaluated at $k=aH$. The last
line is additional contribution which comes from the slow-roll
expansion. The commutative contribution up to second order is
given by \beq \label{n1} n_{s}(k) = 1 - 4\epsilon_1 - 2\delta_1 +
(8\alpha-8)\epsilon_1^2 + (10\alpha -6)\epsilon_1\delta_1 -
2\alpha\delta_1^2 + 2\alpha\delta_2. \eeq

 Finally the running
spectral index up to third order is given by
\begin{eqnarray}
\frac{d\tilde{n}_s}{d\ln k} =& &\frac{d n_s}{d\ln k}
                                  - 32\mu\epsilon_1(
                                  \epsilon_1-\delta_1)
                                  \\ \nonumber
                             & & - \mu\left\{
                                          32\epsilon_1^3
                                          -(160\alpha+70)\epsilon_1^2\delta_1
                                          +(32\alpha-6)\epsilon_1\delta_1^2
                                          +(32\alpha-14)\epsilon_1\delta_2 \right. \\
                                          \nonumber
                                        & &\left. +4\delta_1^3-6\delta_1\delta_2
                                          +2\delta_3
                                       \right\}.
\end{eqnarray}
The last two lines come from the slow-roll expansion.  Also the
commutative contribution up to third order
\begin{eqnarray}
\frac{d}{d\ln k} n_{s} & = &
-8\epsilon^{2}_{1}-10\epsilon_1\delta_1+2\delta^{2}_{1}-2\delta_2+(32\alpha
-40)\epsilon^{3}_{1}
\\ & & +(62\alpha
-60)\epsilon^{2}_{1}\delta_1+(6\alpha
-4)\epsilon_1\delta^{2}_{1}+(14\alpha
-8)\epsilon_1\delta_2+4\alpha\delta^{3}_{1}-6\alpha\delta_1\delta_2+2\alpha\delta_3.
\nonumber
\end{eqnarray}
Up to now our calculation was  done at the pivot scale of $k=k_*$.
In order to obtain the correct expressions for noncommutative
spacetime inflation, we have to change the pivot scale: $k_*\to
k_{nhc}$. Up to first-order corrections, $k_*\to k_{nhc}$
corresponds to $\alpha \to \alpha_*$ in the above
expressions\cite{STW}.  As an example, we choose the power-law
inflation like $a(t)\sim t^p$ whose potential is given by \beq
V(\phi)=V_0 \exp\Big(-\sqrt{\fr{2}{p}}\phi\Big).\eeq Thus
slow-roll parameters are determined  by \beq \label{pislp}
\epsilon_1=\fr{1}{p},~\delta_1=-\fr{1}{p},~
\delta_2=2\delta_1^2=\fr{2}{p^2},~\delta_3=6\delta_1^3=-\fr{6}{p^3}.\eeq
Then the noncommutative power spectrum takes the form \beq
\label{pips}\tilde{P}^{PI,1st}_{R_c}(k)  =
P^{PI,1st}_{R_c}(k)-\frac{\mu H^4}{(2\pi)^2\dot\phi^2} \left\{
       4  +\fr{12(2\alpha_*-1)}{p}
\right\}, \eeq where \beq
 \label{2ndpss}
P^{PI,1st}_{R_c}(k) = \frac{H^4}{(2\pi)^2\dot\phi^2} \left\{ 1
+2(\alpha_*-1)\fr{1}{p} \right\}. \nonumber \eeq The
noncommutative spectral index can be easily calculated up to
second order
\begin{equation} \label{pisi}
\tilde{n}^{PI}_{s}(k) = n^{PI}_{s}(k)+\mu \left\{\fr{16}{p} +
\fr{64\alpha_*}{p^2}\right\},
\end{equation}
where \beq  n^{PI}_{s}(k)=1-\fr{2}{p}-\fr{2}{p^2}. \eeq Finally
the running spectral index is found to be \beq
\label{pirsi}\frac{d\tilde{n}^{PI}_s}{d\ln k} \simeq \frac{d
n^{PI}_s}{d\ln k}
                                  -\mu \left\{\fr{64}{p^2} +\fr{8(32\alpha_*+8)}{p^3}
                                  \right\},
                                  \eeq
 where the commutative contribution is zero up to $1/p^3$,
\beq \frac{d n^{PI}_s}{d\ln k}=0. \eeq Comparing the above
expressions with those of a constant $\mu$ in Ref.\cite{HM3}, we
find additional corrections in equations (\ref{pips}),
(\ref{pisi}), and (\ref{pirsi}). These are replacement of $\alpha
\to \alpha_*$   from the change of pivot scale and additional
contributions from a  noncommutative parameter with $\mu\not=$
constant.

\section{Discussion}
We study the role of the noncommutative parameter in inflation.
First of all we could  consider it as a zeroth order slow-roll
parameter. In the case of the power-law inflation with
$\epsilon_1$=constant \cite{HM3}, one has
$\mu=(k/k_c)^{-4/p}=e^{-\fr{4}{p} \ln(k/k_c)}$. Considering that
$p$ is large, one has an undetermined series  with respect to $p$
in compared with slow-roll parameters in Eq. (\ref{pislp}). Also,
from the last term in Eq.(\ref{zzz}), we have
$\tilde{\eta}=(1+\mu)\eta$. This  means that
 $\mu$ is a dimensionless parameter, which expands the conformal time
 scale in noncommutative spacetime slightly.  Further,  introducing
$k\sim aH$ at horizon crossing,  we have a relation of $\mu \sim
H^4$.   Hence it is no doubt that $\mu$ is regarded as a zeroth
order slow-roll parameter\footnote{For a better notation, we may
use $\mu_0$ instead of $\mu$.}.

 In order to calculate the
noncommutative power spectrum correctly, we have to consider two
important things. The first is to note that the noncommutative
parameter is a variable one which satisfies $\dot{\mu}=-4H
\epsilon_1 \mu$. Then the derivative of $\mu$ with respect to time
leads to increase one order in the slow-roll expansion. In the
cases of $\mu=0$ and  $\mu=$ constant, we have gotten the same
power spectrum up to first order when using the slow-roll
approximation and slow-roll expansion. However, in the case of
$\mu \not=$ constant, two results are different even for
first-order corrections.  The second is to note that the pivot
scale $k=k_{nhc}$ of a noncommutative spacetime inflation is
different from $k=k_*$ of commutative case. There exists  a delay
of the time when fluctuation modes cross outside the horizon in
noncommutative spacetime. We can take it into account   by making
a change of $\alpha \to \alpha_*$. On the other hand, the
slow-roll expansion is suitable for $k=k_*$ and higher order
calculations. Hence, in order to obtain the noncommutative power
spectrum up to first order, we simply replace $\alpha$ by
$\alpha_*$ in whole commutative expressions at $k=k_*$.

As is shown in Eq. (\ref{pips}), choosing the power-law inflation,
we confirm that the presence of $\mu$ (noncommutative spacetime
effect) always suppress the power spectrum. Especially, a case of
$\mu \not=$constant  suppresses more the power spectrum than that
of a $\mu$=constant case.

\subsection*{Acknowledgements}
This  was supported in part by  KOSEF, Project No.
R02-2002-000-00028-0.

\end{document}